\documentstyle[twocolumn,aps,prb,epsfig,floats]{revtex}
\input epsf
\draft

\begin{document}
\twocolumn[\hsize\textwidth\columnwidth\hsize\csname@twocolumnfalse\endcsname

\title{Aging, rejuvenation and memory effects in Ising and Heisenberg
spin glasses\\ }

\author{V. Dupuis${}^{(1)}$, E. Vincent${}^{(1)}$,
J.-P. Bouchaud${}^{(1)}$, J.Hammann${}^{(1)}$, A. Ito${}^{(2)}$,
H. Aruga Katori${}^{(3)}$}
 
\address{${}^{(1)}$Service de Physique de l'Etat Condens\'e, CEA
Saclay,\\ \it 91191 Gif sur Yvette Cedex, France \\ ${}^{(2)}$Muon
Science Laboratory and ${}^{(3)}$Magnetic Materials Laboratory, \it
RIKEN,\\ Hirosawa 2-1, Wako, Saitama 351-0198, Japan}

\maketitle

\begin{abstract}
We have compared aging phenomena in the $Fe_{0.5}Mn_{0.5}TiO_3$ Ising
spin glass and in the $CdCr_{1.7}In_{0.3}S_4$ Heisenberg-like spin
glass by means of low-frequency {\em ac} susceptibility measurements.
At {\em constant} temperature, aging obeys the same `$\omega t$
scaling' in both samples as in other systems.  Investigating the
effect of temperature variations, we find that the Ising sample
exhibits {\em rejuvenation and memory} effects which are {\em
qualitatively} similar to those found in other spin glasses,
indicating that the existence of these phenomena does not depend on
the dimensionality of the spins.  However, systematic temperature
cycling experiments on both samples show important {\em quantitative}
differences.  In the Ising sample, the contribution of aging at low
temperature to aging at a slightly higher temperature is much {\em
larger} than expected from thermal slowing down. This is at variance
with the behaviour observed until now in other spin glasses, which
show the opposite trend of a free-energy barrier growth as the
temperature is decreased.  We discuss these results in terms of a
strongly renormalized microscopic attempt time for thermal activation,
and estimate the corresponding values of the barrier exponent $\psi$
introduced in the scaling theories.
\end{abstract}

\pacs{PACS Numbers~: 75.50.Lk Spin glasses and other random magnets, 
75.10.Nr Spin-glass and other random models, 75.40.Gb Dynamic properties}

\twocolumn]\narrowtext

\section{Introduction}
Spin glasses, an intriguing class of disordered and frustrated
magnetic systems, have been the focus of many efforts for some years.
Below their glass transition temperature, they are out of equilibrium
on all experimental time scales, their dynamics is slow, and {\it
non-stationary}: it is the {\it aging} phenomenon, widely studied
experimentally (see various references in \cite{Sitges,Uppsala}),
theoretically \cite{CuKu,traptree,domains,Review} and numerically
\cite{numrecent}.  Still, the non-trivial dependence of aging
phenomena on temperature changes (`rejuvenation and memory effects'
\cite{memchaos}) is yet poorly understood \cite{Miyashita}, and their
interpretation in terms of the slow growth of `spin-glass ordered
domains' remains unclear \cite{domains,memchaos,bray,Yoshino}.

Until now, rejuvenation and memory effects have only been studied in
samples of weak anisotropy, with Heisenberg-like spins
\cite{Sitges,Uppsala,memchaos,hierarki,barriergrowth}.  The comparison
of these experiments with the numerical simulations of the
Edwards-Anderson model (Ising spins), which is presently the subject
of intensive efforts \cite{numrecent}, may thus not be fully relevant.
Therefore, it is of particular interest to perform experiments on {\it
Ising} compounds along the same procedures as in previous
investigations.

In this paper, we present new {\it ac} susceptibility measurements on
the $Fe_{0.5}Mn_{0.5}\-TiO_3$ spin glass, which is a
well-characterized Ising system \cite{FMTOgeneral}.  Previous studies
\cite{FMTOaging} have shown that some rejuvenation is visible upon
decreasing the temperature.  But memory effects, and their interplay
with rejuvenation, have not yet been investigated. Here, we want to
address the questions of a possible universality of rejuvenation and
memory phenomena in Ising and Heisenberg spin glasses, and of their
quantitative comparison in various systems.

In the Ising sample, we could check that aging at constant temperature
follows the usual $\omega t$ scaling (in agreement with previous dc
experiments \cite{FMTOaging}). We find that {\it rejuvenation and
memory} effects are also present, although more `spread out' in
temperature than in other spin glasses \cite{memchaos,Nordblad}. For a
quantitative comparison, we have taken similar new data on the
$CdCr_{1.7}In_{0.3}S_4$ Heisenberg-like\cite{foot1} spin glass
\cite{Nogues,Sitges}, which allow us to detect and characterize marked
differences.

In the spin glasses explored so far, aging experiments in which a
small temperature cycle is applied from $T$ to $T- \Delta T$ and back
to $T$ have shown that the contribution on aging of the $T- \Delta T$
part of the cycle decreases extremely rapidly for increasing $\Delta
T$ \cite{hierarki,Sitges}.  This drastic slowing down can be seen, in
terms of thermal activation, as an {\it increase} of the free-energy
barrier heights as the temperature is decreased
\cite{barriergrowth}. On the contrary, in the present Ising sample, we
find that this separation of time scales between $T$ and $T-\Delta T$
is {\it weaker} than expected from thermal slowing down; in other
words, this would imply the unlikely scenario of a barrier height {\it
decrease} upon cooling.

We propose to interpret this result as a sign that the elementary
attempt time involved in thermal fluctuations is {\it much longer}
than the paramagnetic time of $\sim 10^{-12}$s, pointing out the
relevance of critical fluctuations below $T_g$. This same picture can
also be applied to the results from the other (Heisenberg-like)
sample, and in Section 5 we discuss all results in terms of the
barrier exponent $\psi$ introduced in scaling theories \cite{domains}.

\section{Basic facts: aging at constant temperature}
The $Fe_{0.5}Mn_{0.5}TiO_3$ compound has been characterized as a
typical Ising spin glass system \cite{FMTOgeneral}. The thermal
variation of the susceptibility, when measured in a field {\it
parallel} to the hexagonal c-axis of the ilmenite-type crystalline
structure, shows a sharp cusp at $T_g=20.7K$ which is absent in the
{\it perpendicular} direction. The Ising character is well evidenced
by the very small ratio of perpendicular to parallel magnetization (of
about $7\%$)\cite{FMTOgeneral}.  The sample used in this study is a
single crystal with the shape of a rectangular parallelepiped (2x2x5
$mm^3$), having its long axis parallel to the c-axis. The measurements
were made along this axis, using a commercial SQUID magnetometer
(Cryogenic Ltd, U.K.) with a $1.7\ Oe$ peak value of the {\it ac}
field.

Previous {\it dc} experiments on $Fe_{0.5}Mn_{0.5}TiO_3$
\cite{FMTOaging} have shown that the relaxation rate of the Zero Field
Cooled magnetisation shows a broad maximum in the region $ln\ t \sim
ln\ t_w$, compatible with a $t/t_w$ scaling. Here we study the
frequency ($\omega$) dependence of aging in the {\it ac}
susceptibility, in order to test the validity of $\omega t$ scaling
(equivalent to $t/t_w$ scaling in {\it dc} experiments). We quench the
sample from above $T_g$ to a given temperature $T<T_g$ (here $19K$ and $15K$), and
measure the subsequent relaxation of the {\it ac} susceptibility at
this constant temperature $T$ as a function of
time $t$. Each point consists in the successive measurements of $8$
frequencies in the range $0.04Hz-8Hz$. As in the rest of the paper, we
use the data in the paramagnetic regime (assuming $\chi''$ = 0) for
checking and correcting slight frequency-dependent phase shifts in the
detection setup.

We present in Fig.1 a scaling plot for the imaginary part $\chi''$ of
the {\it ac} susceptibility $\chi''(\omega,t)$ at $T_0=19K$. The
curves have been shifted on the vertical scale in such a way that they
superimpose as a function of $\omega t$. This (usual) procedure
accounts for the stationary part $\chi''_0(\omega)$ of the
susceptibility (i.e. its limit as $t \rightarrow \infty$), which can
on rather general grounds \cite{Sitges} be subtracted in order to
separate the aging part. The aging part can be seen to satisfactorily
obey an $\omega t$ scaling. This also holds for the real part $\chi'$
at $19K$ (inset of Fig.1), and as well for both $\chi'$ and $\chi''$
at $15K$.  We have fitted the curves of Fig.1 (and the corresponding
ones at $15K$) to a power law 
\begin{equation}
\chi''(\omega,t)$- $\chi''_0(\omega)$=$A\, (\omega t)^{-b}\ \ \ . 
\label{chi''fit}
\end{equation}
The best fit yields $b=0.14\pm 0.03$ for both
temperatures. For a direct comparison, we have taken new data on the
$CdCr_{1.7}In_{0.3}S_4$ thiospinel compound ($T_g=16.7K$) at the same
reduced temperatures. They yield, by the same analysis, $b=0.18\pm
0.03$. Let us note that, since the value of $b$ is not strongly
constrained (as indicated by the error bars), we cannot conclude that
a significant difference in $b$ exists between both systems.  However,
the aging magnitude $A$ is better determined, and shows a clear
temperature dependence in the Ising system ($20\%$ decrease from $19K$
to $15K$, while it is constant in the thiospinel in this same range of
temperature).

We have also analyzed the stationary susceptibility $\chi_0(\omega)$,
as determined from both the superposition of the relaxation curves at
different frequencies (Fig.1) and the non-linear fit of Eq.\ref{chi''fit}.  
The stationary susceptibility of 
the Ising sample fits well to a power law $\chi''_0
\propto\omega^{\alpha}$ with a small exponent $\alpha\sim 0.1$, in the
same range ($0.02-0.10$, see \cite{chisecpowerlaw}) as commonly
observed in other (Heisenberg-like) spin glasses. We have checked that
the real part $\chi'_0$ of the susceptibility obeys the corresponding
law $\chi'_0= \chi'_{stat}-{\mathcal A}\ \omega^{\alpha}$ with the
same exponent values.

Thus, from our {\it ac} measurements, the low-frequency behaviour of
this Ising sample at {\it constant} temperature is similar to that of
Heisenberg-like systems.

\begin{figure}[htbp]
\begin{center}
   \leavevmode
   \epsfysize=7cm\epsfbox{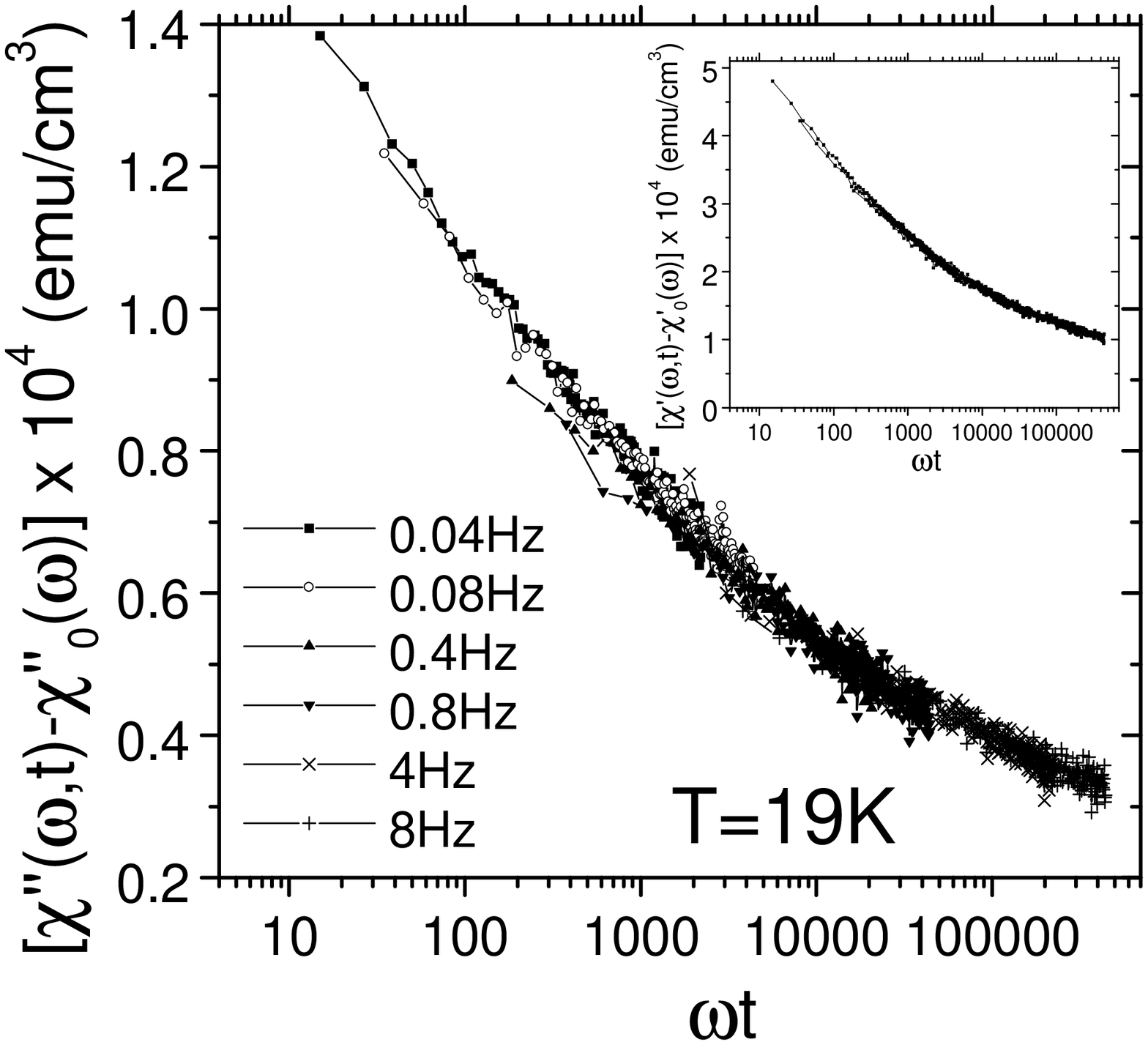}
\end{center}
\caption{\label{scaling}
Out of phase susceptibility $\chi''(\omega,t)$ vs. $\omega t$ for
the $Fe_{0.5}Mn_{0.5}TiO_3$ sample. The six curves have been
vertically shifted (see text). $t$ is the total time elapsed since
the quench (age). The inset shows the same scaling for the in
phase susceptibility $\chi'$. }
\end{figure}

\section{Rejuvenation  and memory effects in the Ising sample}
Temperature cycling protocols have revealed the `rejuvenation and
memory' effects in several (Heisenberg-like) spin glasses
\cite{hierarki,memchaos,Sitges,Nordblad}. We have applied the same
procedures to the Ising sample.  In a first experiment (Fig.2), the
sample is quenched from above $T_g$ to $T_1=18K$ and the relaxation of
the {\it ac} susceptibility at frequency $0.4Hz$ is recorded as a
function of time. After a time $t_1=4000s$, the sample is subjected to
a negative temperature cycle of duration $t_2=4000s$ and amplitude
$2K$. Cooling from $T_1=18K$ to $T_2=16K$ induces a strong restart of
aging (see Fig. 2): it is the so-called {\it rejuvenation} effect. The
renewed relaxation seems to take place independently of the previous
equilibration stage at $T_1$. But when the temperature is turned back
to $T_1$ after the cycle, we see that the two relaxations at $18K$ are
in continuation of each other (inset of Fig. 2). The system has
actually kept a {\it memory} of the state reached before the cycle,
and the strong relaxation at $16K$ had no effect on the relaxation at
$18K$.

\begin{figure}[htbp]
\begin{center}
   \leavevmode
   \epsfysize=7cm\epsfbox{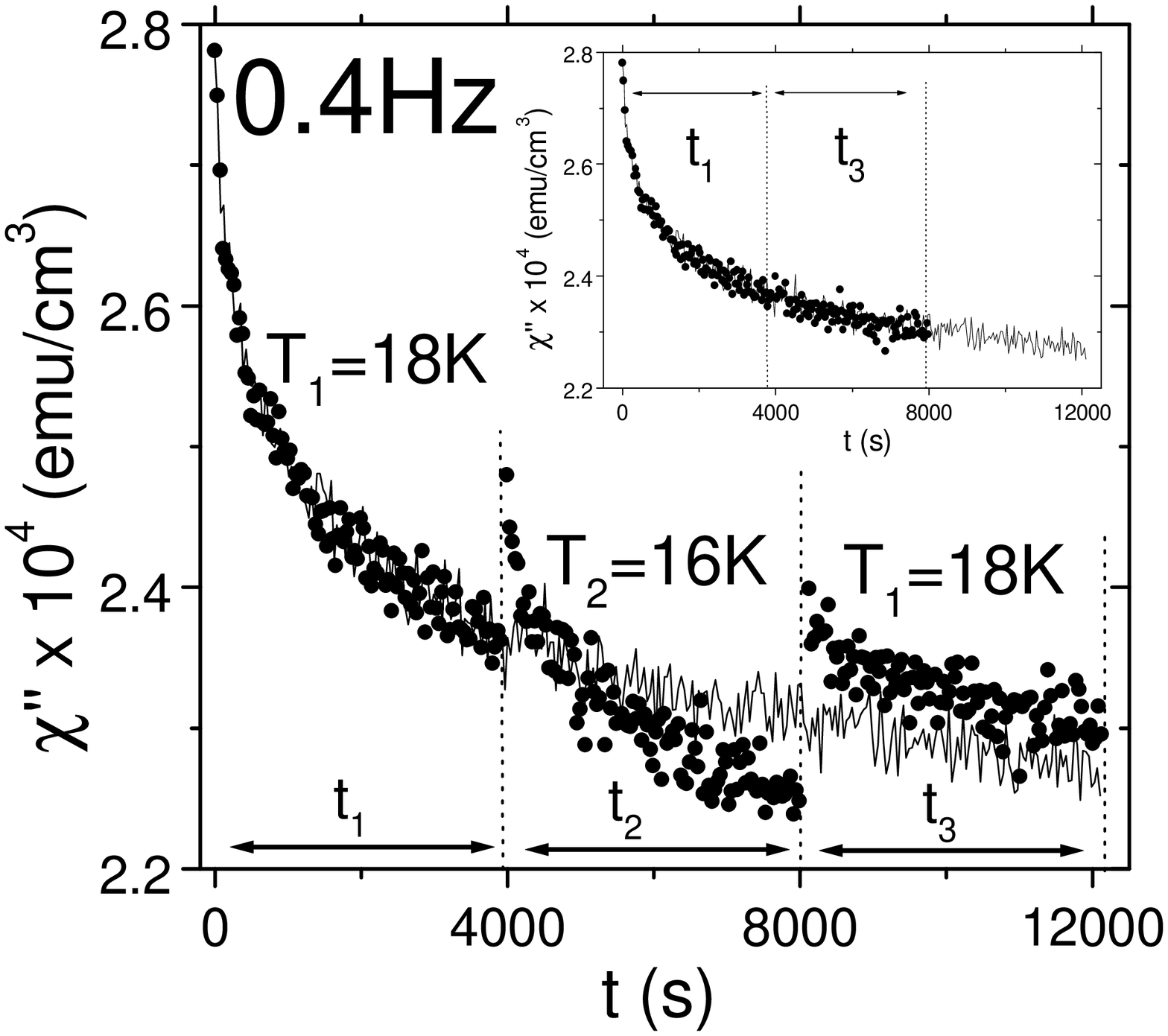}
\end{center}
\caption{\label{memtot}
Out of phase susceptibility $\chi''(\omega,t)$ of the
$Fe_{0.5}Mn_{0.5}TiO_3$ sample during a negative temperature
cycle. The inset shows that, despite the strong relaxation at
$16K$ (rejuvenation effect), both parts at $18K$ are in
continuation of each other (memory effect). }
\end{figure}

In a second experiment, we apply a procedure in which rejuvenation and
memory effects can be visualized as `dips' in the thermal variation of
$\chi''$ \cite{memchaos,Nordblad}.  Starting from the paramagnetic
phase, the {\it ac} susceptibility is measured (at 0.04, 0.4 and 4 Hz
in one run) while continuously cooling down to $3K$ at a rate of 0.001
K/s, except for 3 stops of 7h at $19K$, $15K$ and $10K$, during which
we let the sample age at constant temperature. Then the temperature is
continuously raised back to $30K$ at the same rate. In a separate run,
we determine {\it reference curves} (solid lines in Fig.3)
corresponding to the same protocol {\it without stops}.

At each aging temperature, $\chi''$ slowly relaxes downwards with
time, as seen in Fig.3.  The lower the frequency, the stronger
the relaxation, in agreement with the $\omega t$ scaling (previous
section).  Rejuvenation effects are clearly observed at the two higher
aging temperatures of $19K$ and $15K$; when cooling is resumed after
aging, $\chi''$ rises up and merges with the reference curve (although
this reference curve itself is out of equilibrium).  The other
important feature is seen on the $\chi''$ curve measured upon
re-heating continuously from 3K. When approaching $15K$ and $19K$,
$\chi''$ departs from the reference curve and traces back a dip which
displays the memory of the past relaxation at those
temperatures. The rejuvenation and memory effects obtained in this
Ising sample are {\it qualitatively} similar to those observed in
other spin glasses \cite{memchaos,Nordblad}.

\begin{figure}[htbp]
\begin{center}
   \leavevmode
   \epsfysize=7cm\epsfbox{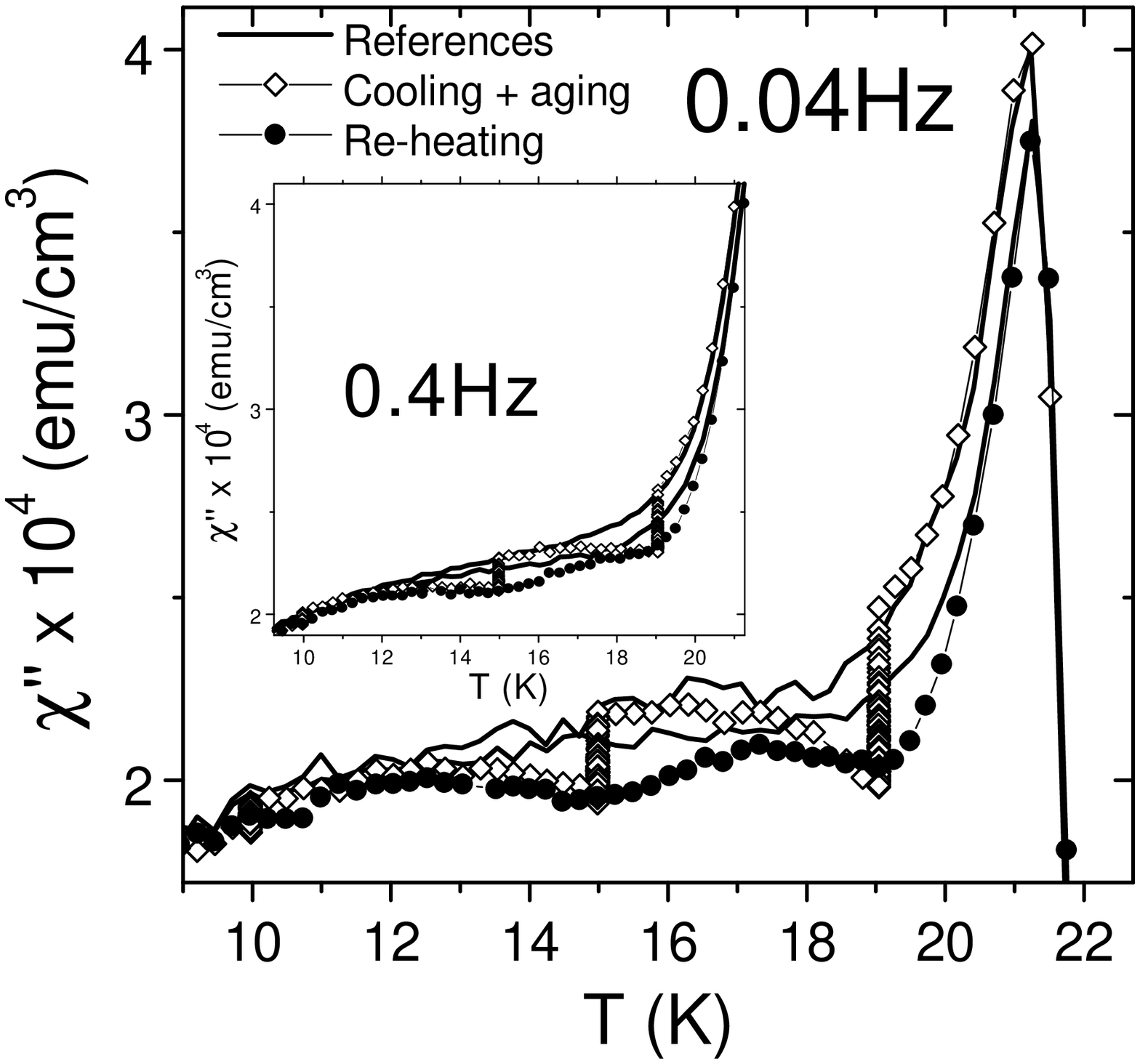}
\end{center}
\caption{\label{memorej} Out of phase susceptibility
$\chi''(\omega,t)$ vs. temperature at $\omega=0.04Hz$ for the
$Fe_{0.5}Mn_{0.5}TiO_3$ sample. The solid lines show a reference
behaviour for continuous cooling and re-heating (the re-heating curve
is slightly lower than the cooling curve). Open diamonds: when cooling
is stopped at 19K, 15K and 10K (resp. $0.9$, $0.7$ and $0.5T_g$),
$\chi''$ relaxes due to aging (very weakly at 10K) but when cooling
resumes $\chi''$ merges with the reference curve. Solid circles: when
re-heating after cooling with stops for aging, the memory of aging is
retrieved. The inset shows the same results for $\omega=0.04Hz$. }
\end{figure}

However, in the experiment of Fig.3 with successive agings stops at
$0.9T_g$, $0.7T_g$ and $0.5T_g$, the observed relaxation decreases
significantly from one stop to another (by a factor 2.5 between
$0.9T_g$ and $0.7T_g$), and becomes undistinguishable at
$0.5T_g$. This feature makes a difference with the
$CdCr_{1.7}In_{0.3}S_4$ spin glass\cite{memchaos}, and may indicate
some `accumulation' of aging, which would in consequence progressively
die out during a long cooling procedure with multiple stops.

In fact, the comparison of the $\chi''$ relaxations at $0.7T_g$ after
a slow cooling ($0.001K/s$ as above) and a direct quench does not show
any difference between both samples: in the slow cooling case, the
relaxation is for both samples $\sim 70 \%$ of that obtained after the
direct quench (of course, rejuvenation effects do not exclude that the
very last part of the cooling procedure induces a weak but non-zero
cooling rate effect\cite{memchaos}).  But, in a multiple stop
procedure like the one in Fig.3, we could check that in this Ising
sample the relaxation at the {\it lower} temperature is {\it more}
influenced by aging at the higher temperature than in the thiospinel.
The relaxation at $0.7T_g$ (second stop) only amounts to $\sim 25 \%$
of that obtained after a direct quench, whereas this proportion is of
$\sim 50 \% $ for the thiospinel in the same conditions.

 In other words, aging processes in this Ising sample are more
sensitive to the thermal history (cooling rate effect) than in other
spin glasses, as already suggested by the spreading in temperature of
the dips in Fig.3.

\section{Quantitative experiments: thermal slowing down and barrier growth}
The procedure of negative temperature cycling (like presented in
Fig.2) is `simpler' than the continuous cooling/re-heating procedure
of Fig.3, and can yield more direct informations on the influence of
time and temperature on the rejuvenation and memory effects. We now
discuss a series of such measurements on both samples, with {\it
small} temperature intervals.  After quenching from above $T_g$ to
$T<T_g$ at time zero, the temperature $T$ is kept constant for a time
$t_1=7700s$, after which it is lowered to $T-\Delta T$ for a time
$t_2=23650s$, and finally raised back to $T$ for a time $t_3= 6000s$.

The relaxation of the {\it ac} susceptibility at frequency $\omega =
0.1Hz$ is recorded during the whole procedure.  As working
temperatures, we chose equivalent values in units of $T_g$ for both
samples: 
\begin{itemize}

\item $T=18K=0.87T_g$ and $T=15K=0.72T_g$ for
$Fe_{0.5}Mn_{0.5}TiO_3$,

\item $T=14K=0.84T_g$ and $T=12K=0.72T_g$ for
$CdCr_{1.7}In_{0.3}S_4$.  

\end{itemize}
On the other hand, the temperature
intervals have been adjusted in order to produce about the same effect
in both samples, therefore they are different, even in units of $T_g$:
$\Delta T$=0.25, 0.5, 0.75 and 1K for $Fe_{0.5}Mn_{0.5}TiO_3$, and
$\Delta$T=0.1, 0.2, 0.3 and 0.4K for $CdCr_{1.7}In_{0.3}S_4$.

An example experiment is presented in Fig.4 for the Ising sample at
$T=15K$. The effect of a $\Delta T=1K$ negative cycling is compared
with an isothermal reference curve at $T=15K$.  For this small
temperature interval, {\it no jump} (such as the one seen in Fig.2) in
the relaxation signs up any important rejuvenation effect as the
temperature is decreased, whereas a {\it small jump} of $\chi''$ is
visible for the thiospinel sample in the same
conditions.  When the
temperature is raised back to $T$, $\chi''$ rises up to a value which
is higher than the isothermal reference at $T$.  This final relaxation
can be superposed with the isothermal reference by a shift backwards
of a time $t_2 -t_{eff}$ with $t_{eff}=4500s$ (see inset of Fig.4).
The effective time $t_{eff}$ accounts for the effect at $T$ of the
processes which occurred at $T-\Delta T$ during $t_2$.  After
$t_1+t_2$, the state of aging can be considered as an additive
combination of aging at $T$ before the cycle, which has hence been
memorized, with a contribution $t_{eff}$ from aging at $T-\Delta T$.
It is interesting to compare this result with a simple estimate of the
effect of thermal slowing down.  Considering simply that aging
involves thermally activated jumps over {\it fixed} free-energy
barriers, the effective time $t_{eff}^{TA}$ at $T$ which corresponds
to $t_2$ at $T-\Delta T$ reads:

\begin{equation}
(T-\Delta
T)\log (\frac{t_2}{\tau_0})=T \log(\frac{t_{eff}^{TA}}{\tau_0})
\label{Arrh}
\end{equation}

\noindent where $\tau_0$ is an attempt time, which we choose for now 
as a microscopic spin flip time, say $\tau_0=10^{-12}s$.
Surprisingly, one gets $t_{eff}^{TA}=1915s$, a value smaller than
\hbox{$t_{eff}=4500s$}. That is, aging at the lower temperature has a
{\it larger} contribution than expected from thermal activation
over {\it fixed height} barriers.

\begin{figure}[htbp]
\begin{center}
\epsfysize=7cm\epsfbox{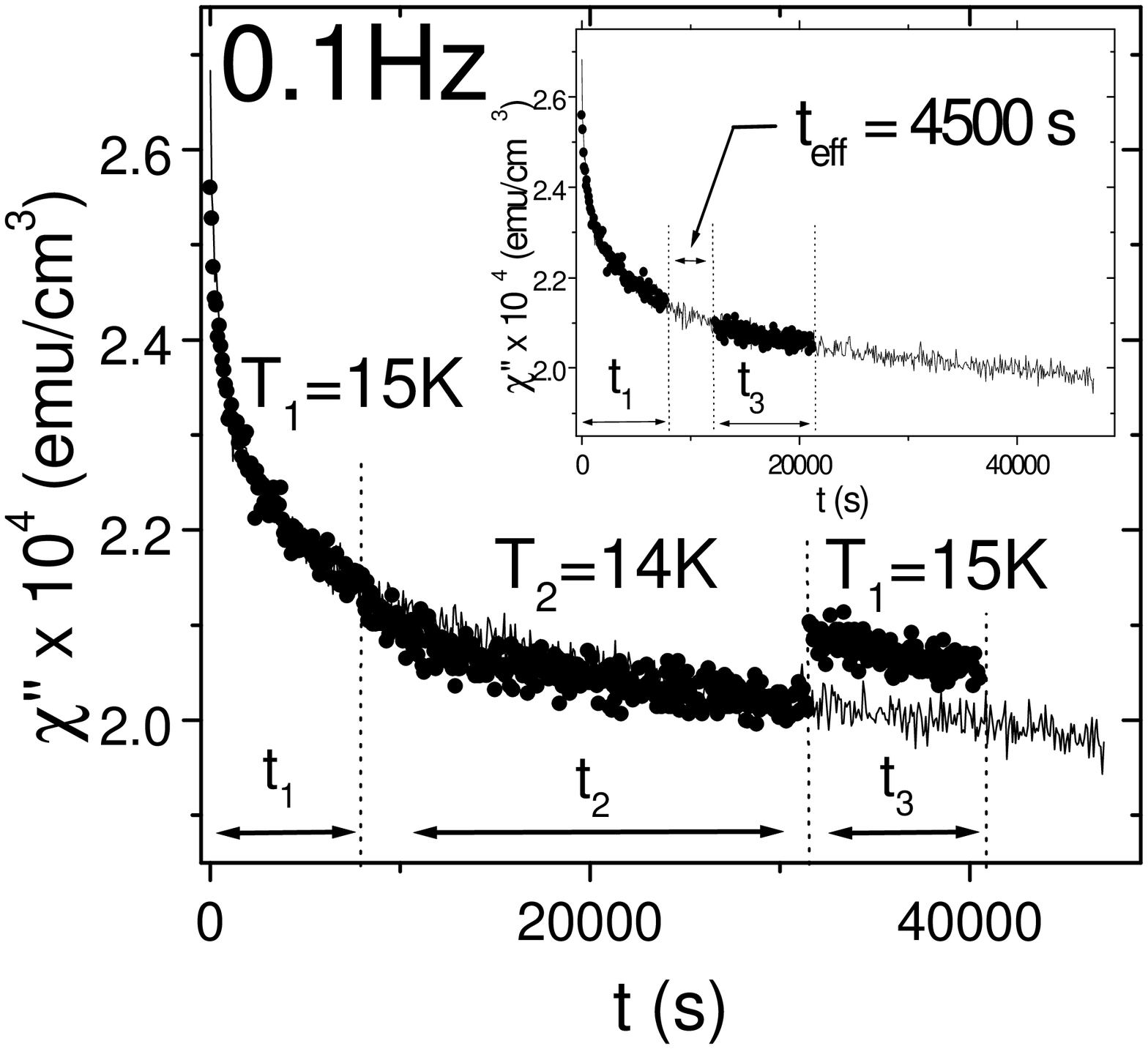}
\end{center}
\caption{\label{cycling}
Effect of a negative temperature cycling ($\Delta T$=-1K) on the
time dependence of $\chi''(\omega = 0.1Hz)$ in the
$Fe_{0.5}Mn_{0.5}TiO_3$ sample. The inset shows a plot of the data
points recorded at $T=15K$ during time $t_1$ before the cycle and
$t_3$ after the cycle. The points after the cycle have been
superposed on a standard isothermal relaxation curve by shifting
the time scale. The decay during $t_3$ is a continuation of the
initial decay with an effective time interval \hbox{$t_{eff}$ = 4500s}. }
\end{figure}

\begin{figure}[htbp]
\begin{center}
\epsfysize=7cm\epsfbox{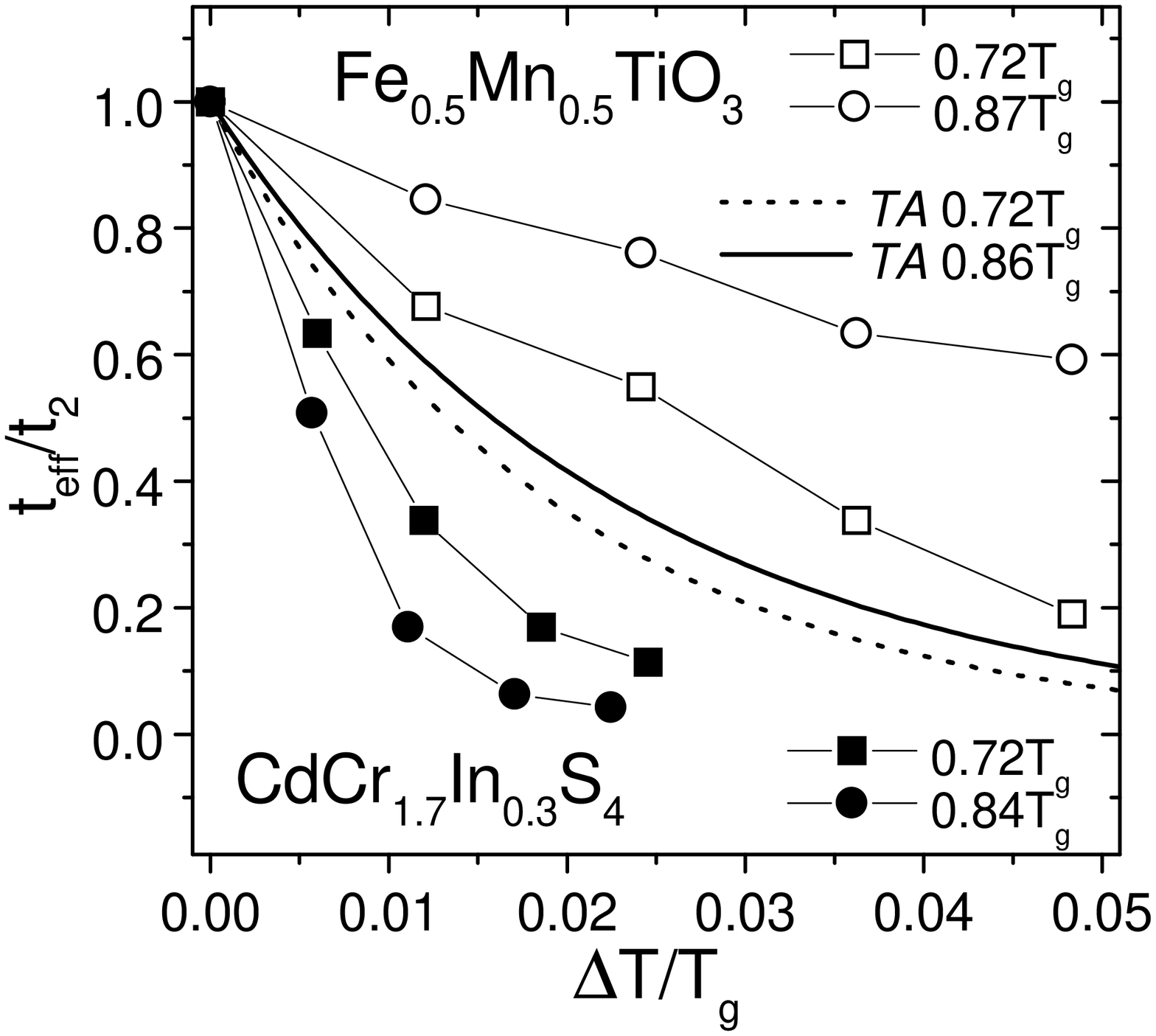}
\end{center}
\caption{\label{tweff}
Effective waiting times extracted from negative temperature cycling
experiments on the $Fe_{0.5}Mn_{0.5}TiO_3$ and
$CdCr_{1.7}In_{0.3}S_4$ samples, normalized to the duration of the
cycle.
For a comparison, the lines ({\it TA}) show the estimate for a simple thermal
slowing down scenario (Eq.(\ref{Arrh}) with $\tau_0=10^{-12}s$).}
\end{figure}

This result is in sharp contrast with those obtained for the
thiospinel sample and other spin glasses, in which the effective
contribution of aging at a lower temperature is always smaller than
expected from simple thermal slowing down, a general result that has
been discussed in terms of a growth of free-energy barriers as the
temperature is lowered \cite{hierarki,barriergrowth}.

The whole set of results on both samples is summarized in Fig.5, which
displays the measured values of the effective times $t_{eff}(T, \Delta
T)$, in comparison with the calculated estimates for thermal
activation over {\it fixed height} barriers (Arrh\'enius law,
Eq.(\ref{Arrh}), with $\tau_0=10^{-12}$s). Clearly, two different
behaviours are obtained: for the thiospinel (Heisenberg-like) sample,
we always have $t_{eff} < t_{eff}^{TA}$, whereas $t_{eff} >
t_{eff}^{TA}$ for the Ising one.

\section{Physical interpretation: a temperature dependent length scale}

Our result that, in the Ising sample, the contribution of aging slows
down less than expected when the temperature is lowered is, at first
sight, puzzling.  However, this conclusion is based on the assumption
that the relevant elementary time scale is the paramagnetic spin-flip
time $\tau_0 \sim 10^{-12}$s.  In Fig.5, the position of the curves
calculated for thermal activation depends on the choice of the attempt
time $\tau_0$; they would move upwards for larger values of $\tau_0$.
Another way to formulate this is proposed in Fig. 6, where we have
plotted $\log [t_{eff}/t_2^{T-\Delta T/T}]$ as a function of $\Delta
T/T$. If barriers of the same height are crossed during the
experiments at $T$ and $T-\Delta T$ with an attempt time $\tau_0$, one
should observe a straight line of slope $\log \tau_0$
(Eq.\ref{Arrh}). One can see very clearly that the Ising data
corresponds to an effective attempt time $\tau_0^{eff}$ much larger
than $10^{-12}$ seconds, whereas the $CdCr_{1.7}In_{0.3}S_4$
Heisenberg-like spin glass favours $\tau_0^{eff} \ll 10^{-12}$ s.  A
comparison can also be made with a third example, the $Ag:Mn\ 2.7\%$
spin glass, in which the anisotropy should still be weaker.  In this
sample, very detailed studies on the effect of tiny temperature
variations have been performed in {\it dc} experiments
\cite{barriergrowth}, yielding a precise characterization of the
free-energy barrier growth as the temperature is lowered. Fig.6 shows
that the corresponding effective attempt times $\tau_0^{eff}$ are
still shorter.

\begin{figure}[t]
\begin{center}
\epsfysize=7.2cm\epsfbox{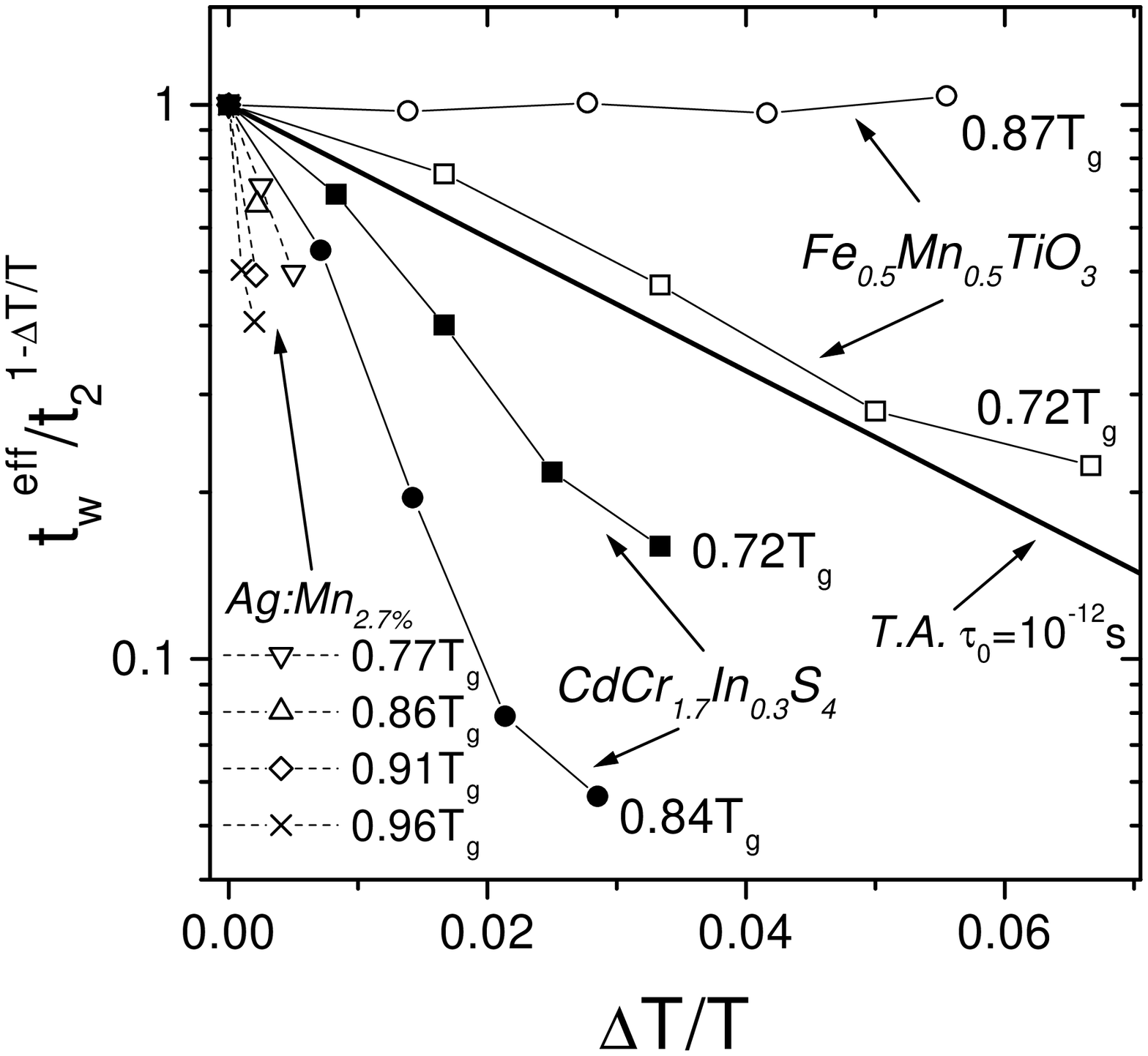}
\end{center}
\caption{\label{tweff_univ}
Effective waiting times extracted from the experiments, plotted in
such a way that thermal activation over fixed height barriers
(Eq.\ref{Arrh}) corresponds to a straight line of slope $\log \tau_0$
($\tau_0$ in s). 
Thus, the average slope of each set of data shows an
estimate of $\log \tau_0^{eff}$, where $\tau_0^{eff}$ would be an
effective attempt time for fixed height barriers.  In addition to the
present data from the $Fe_{0.5}Mn_{0.5}TiO_3$ and
$CdCr_{1.7}In_{0.3}S_4$ samples (same as in Fig.5), we have plotted
comparable results obtained earlier with the $Ag:Mn\ 2.7\%$ sample
\cite{barriergrowth}. }
\end{figure}

A natural possibility to explain the very large increase of the
attempt time in the Ising sample is the proximity of the spin-glass
transition which leads to critical slowing down. Dynamic critical
scaling above $T_g$ tells us that characteristic times $\tau$ are
diverging like the spin-glass correlation length $\xi$ as a power law
of temperature:

\begin{equation}
\tau \sim \xi^z \sim \left ( \frac{T-T_g}{T_g} \right ) ^{-z \nu}
\end{equation}

By analogy with the ferromagnetic case, in which the correlation
length gives the size of the typical (critical) fluctuations {\it
above} as well as {\it below} the transition with the same exponent,
we shall assume that the attempt time in the spin glass phase is
governed by critical fluctuations below $T_g$. During aging, the
characteristic length $\ell$ over which spins are correlated grows
progressively with time, as identified in experiments on the effect of
field variations on the dynamics \cite{xiOrbach} (with good
quantitative agreement with numerical simulations \cite{numrecent}).
We propose that the attempt time corresponding to the length $\ell$ is
proportional to $\ell^z$, and that the time $\tau(\ell)$ which is
needed to flip the $\ell$-sized cluster of spins is governed by
thermal activation over a barrier $B_T(\ell)$, in such a way that
\cite{bouchaudNew}:

\begin{equation}
\tau(\ell)=\tau_0\ \ell^z \ \exp(\frac{B_T(\ell)}{k_BT})
\label{t(l)}
\end{equation}

\noindent where $\ell$ is in units of the lattice constant. 
The temperature and length dependence of $B_T(\ell)$ can
also be expressed by analogy with the ferromagnetic case, and in the
spirit of the scaling theories \cite{domains}. One can write that
\begin{equation}
B_T(\ell)=\Upsilon(T)\ell^\psi \ \ \ , 
\label{B_T}
\end{equation}
where $\Upsilon(T)$ is a wall
stiffness which should vanish at $T_g$ like in ferromagnets:
\begin{equation}
\Upsilon(T)=\Upsilon_0 T_g[(T_g-T)/T_g]^{\psi \nu}\ \ \ .  
\label{Ups}
\end{equation}
This corresponds
to the assumption that barriers on the scale of the correlation length
$\xi$ are of order $k_B T_g$.

Replacing the attempt time $\tau_0$ by a longer time $\tau_0 \ell^z$
in Eq.(\ref{Arrh}) will indeed qualitatively account for our
experimental results on the Ising sample (weak slope $\log
\tau_0^{eff}$ of the $Fe_{0.5}Mn_{0.5}TiO_3$ data in Fig.6). The
$\tau_0 \ell^z$ term is dominant in Eq.\ref{t(l)} provided one is in a
regime where $B_T(\ell) \leq k_BT$, i.e. for temperatures sufficiently
close to $T_g$. For lower temperatures, one may expect the growth of
the barriers when the temperature is decreased (i.e the $\Upsilon(T)$
variation, Eq.\ref{Ups}) to be the dominant effect, therefore leading
to a much stronger slope $\log \tau_0^{eff}$. Our results indicate
that the Ising sample is dominated by the critical regime, whereas the
Heisenberg-like samples are rapidly dominated by the barriers (see
below the values $z, \nu$ and $\psi$).

We have adjusted our experimental data to Eqs. \ref{t(l)}, \ref{B_T},
\ref{Ups} in the following way.  What comes out from each negative
cycling experiment is the equivalence between an aging time $t_2$ at
$T_2=T_1-\Delta T$ and an effective time $t_{eff}$ at $T_1$ (to be
fully rigorous, the equivalence should also involve aging during $t_1$
at $T_1$, but we checked that this does not affect the results).  In
terms of the characteristic length scales involved, this equivalence
corresponds to $\ell(t_{eff},T_1)=\ell(t_2,T_2)$.  Using
Eq.(\ref{t(l)}), we can numerically extract $\Upsilon_0$ and $\psi$
from the experiments by a best fit procedure \cite{bouchaudNew}, which
allows in most cases to obtain $\ell(t_{eff},T_1)=\ell(t_2,T_2)$
within about $0.5\%$.  In practice, both parameters are strongly
correlated in the fit. We therefore present the results obtained with
a reasonable value $\Upsilon_0=1$. Choosing higher (resp. lower)
values of $\Upsilon_0$ yields lower (resp.  higher) values of $\psi$,
but all qualitative trends remain the same. On the other hand, we have
taken $z\nu$ from results of dynamic critical scaling , and estimated
$\nu$ from the static scaling exponents $\beta$ and $\gamma$, using
$\nu=(2\beta +\gamma )/d\ $ \cite{FMTOgeneral,Nogues,Helene}. Thus we
have $\nu=1.7$ and $z\nu=10.5$ for $Fe_{0.5}Mn_{0.5}TiO_3$, $\nu=1.3$
and $z\nu=7$ for $CdCr_{1.7}In_{0.3}S_4$, and in addition we have used
$\nu=1.4$ and $z\nu=5\ $ \cite{Helene,Tholence} for a comparison with
older results on $Ag:Mn2.7\%$ \cite{barriergrowth}.  Varying slightly
$z\nu$ and $\nu$ does not affect the following conclusions ($\psi$ may
vary by $\pm 0.1$).

For a given temperature $T$, no systematic variation of $\psi$ with
$\Delta T$ can be seen, therefore we have fitted together all $\Delta
T$'s results at each $T$. The results are as follows (see a summary in
Table 1). In the $Fe_{0.5}Mn_{0.5}TiO_3$ Ising sample, we find
$\psi\sim 0.3$ at $0.87T_g$ and $\psi\sim 0.7$ at $0.72T_g$. This
rather small value of $\psi$ is associated to very large values of the
renormalized attempt time: we find that $\tau_0 \ell^z \sim 2500$ s at
$0.87T_g$ and $5$ s at $0.72T_g$. Hence, the thermal activation hardly
plays any role, in particular at $0.87T_g$: the dynamics is primarily
probing the critical regime, in qualitative agreement with the
`anomalous' effect of temperature shifts reported above. This might
also explain why our estimate for $\psi$ differs between the two
temperatures, since the scaling behaviour of \hbox{$B_T(\ell)\sim
\ell^\psi$} implicitly assumes that $\ell \gg \xi$.

On the other hand, the $CdCr_{1.7}In_{0.3}S_4$ thiospinel sample
(Heisenberg-like) shows a markedly different behaviour. Nevertheless,
our theoretical analysis in terms of temperature dependent barriers
and a renormalized attempt time can be applied to this sample. About
the same value of $\psi\simeq 1.1$ is favoured for both temperatures.
The higher value of $\psi$ confirms the stronger separation of the
different aging contributions for different temperatures, as already
emphasized in the previous Section.

We have also re-analyzed the former {\it dc} data of the $Ag:Mn\
2.7\%$ spin glass \cite{barriergrowth}.  In this third case, the above
analysis leads to $\psi \sim 1.3$, indicating a still stronger effect
of temperature (see \cite{bouchaudNew} for a more detailed
discussion), in agreement with the trend observable in Fig.6.

\begin{table} [htbp] 
\caption{Barrier exponent $\psi$, extracted from the negative cycling
experiments using Eq.\ref{t(l)} with $\nu$ and $z$ as listed. 
The first two lines
correspond to the new {\it ac} measurements presented in this paper; the
$Ag:Mn\ 2.7\%$ result has been deduced from previous {\it dc} data 
\cite{barriergrowth}.\label{table1}}
\begin{tabular}{lccc}
 &$\nu$&$z\nu$&$\psi$\\
\tableline
$Fe_{0.5}Mn_{0.5}TiO_3$ & 1.7 & 10.5 & 0.3-0.7 \\ 
$CdCr_{1.7}In_{0.3}S_4$ & 1.3 & 7 & 1.1 \\ 
$Ag:Mn\ 2.7\%$ & 1.4 & 5 & 1.3 \\
\end{tabular}
\end{table}

\section{Conclusion}
We have applied temperature variation protocols in experiments on the
Ising $Fe_{0.5}Mn_{0.5}TiO_3$ spin glass. We find that the aging
phenomena in the Ising sample present {\it rejuvenation and memory}
effects which are qualitatively similar to those already evidenced in
other (Heisenberg-like) spin glasses \cite{Isingpreprint}.

The interest for the question of the `universality' of rejuvenation
and memory phenomena in spin glasses has been recently revived by the
numerous numerical simulations \cite{numrecent} of the
Edwards-Anderson model (Ising spins), in which these effects could not
yet clearly be seen. Our results indicate that the dimensionality of
the spins should not be at the origin of this discrepancy, which we
rather ascribe to the difference in time scales. The magnetization and
susceptibility experiments are performed at macroscopic times which
are up to $\sim 10^{17}$ times the microscopic spin flip time ($\sim
10^{-12}s$), whereas numerics are until now limited to $\sim 10^{5-6}$
steps, exploring a much shorter range of correlation lengths. It is
plausible that, in simulations, these aging lengths are not enough
different at different temperatures to allow the `separation of length
scales' \cite{bouchaudNew} which is necessary to observe rejuvenation
and memory.

When studied in more details, aging in the Ising sample shows a new
behaviour. Firstly, it dies out rather rapidly towards low
temperatures, and even seems completely frozen below $0.5T_g$.
Secondly, negative temperature cycling experiments show that the
separation of time scales in a negative cycle is here much {\it
weaker} than in other spin glasses, and, surprisingly, even weaker
than expected from thermally activated dynamics with a microscopic
attempt time. We interpret this effect as the signature of the
mesoscopic or even macroscopic value of the attempt time for thermal
activation, which we associate with a characteristic time $\tau
\propto \ell^z$ for critical fluctuations. In this framework, we have
analyzed the temperature variation of the free-energy barriers within
scaling hypotheses inspired from the droplet models of the spin glass
phase \cite{domains}, and found that the analysis could be applied
equally well to both the Ising and the Heisenberg samples. The values
of the barrier exponent $\psi$ that we extract from the experiments on
the Ising sample along this line are of the order of $\sim 0.5$. In
comparison, the same experiments on the $CdCr_{1.7}In_{0.3}S_4$
Heisenberg-like spin glass yield $\psi \simeq 1.1$. Using former data
(from {\it dc} experiments \cite{barriergrowth}) on $Ag:Mn$, we found
$\psi \simeq 1.3$.

Of course, the data on both Heisenberg-like samples do not by
themselves imply considering that the attempt time $\tau_0$ for
thermal activation (Eq.(\ref{Arrh})) should be mesoscopic.  Using a
microscopic $\tau_0 \sim 10^{-12} s$, a rapid growth of the barriers
with decreasing temperature was reported in \cite{barriergrowth}, a
result that was interpreted there as a sign that all barriers
eventually diverge in the spin glass phase. Here, we rather propose
that barriers vanish continuously at $T_g$ and are always finite (for
finite $\ell$) for $T < T_g$ (see also \cite{bouchaudNew}). The
experimental data, over a limited time range, cannot really
discriminate these two scenarii, although the latter has a rather
natural physical motivation.

In any case, the central point of the present paper is that, for the
Ising sample, a microscopic $\tau_0$ would correspond to an unphysical
decrease of the barriers at low temperatures ($\psi <0$). Allowing for
a renormalized attempt time, we find that $\psi$ is positive in the
Ising sample, but significantly smaller than in the other
systems. This small value implies a weaker separation of the time and
length scales as the temperature is lowered. This conclusion does not
depend on the details of the theoretical analysis and is clearly
visible in several qualitative aspects of the data (e.g. Fig.5 and
Fig.6).

It is worth noticing that, within the three samples compared here, the
estimated $\psi$ values are found to increase with decreasing spin
anisotropy, at variance with the intuitive expectation that walls made
of anisotropic spins might be softer (i.e. with lower $\psi$). This
may be related to results from mesoscopic electrical noise
measurements \cite{Meyer}, which show that the temperature dependence
of stationary dynamics is stronger in $CuMn$ spin glasses (equivalent
to barrier growth for decreasing temperature) than in $AuFe$
(Arrhenius behaviour) compounds of much stronger anisotropy. On the
other hand, a systematic dependence of the static critical exponents
at the spin-glass transition as a function of anisotropy has been
recently noted \cite{Campbell}. The question of a different nature of
the spin-glass phase for Ising and Heisenberg spins is a long-standing
issue, and must be considered in the light of the present and
forthcoming experimental and numerical results.

\vskip 1.0cm
\noindent
{\it Acknowledgements}

\noindent
We wish to thank M. Alba, E. Bertin, D. H\'erisson, M. Ocio,
H. Takayama and H. Yoshino for stimulating discussions
during this work.


\vskip-12pt
\begin{thebibliography}{99}


\bibitem{Sitges}E. Vincent, J. Hammann, M. Ocio, J.-P. Bouchaud,
L.F. Cugliandolo, in {\it Complex Behaviour of Glassy Systems},
Springer Verlag Lecture Notes in Physics Vol.492, M. Rubi Editor,
1997, pp.184-219.

\bibitem{Uppsala}P. Nordblad and P. Svedlindh, in `Spin-glasses and Random
Fields', pp.1-28, A. P. Young editor (World Scientific, 1998).

\bibitem{CuKu} L.F. Cugliandolo and J. Kurchan, {\it J. Phys. A:
Math. Gen.} {\bf 27}, 5749 (1994), and {\it Phys. Rev. B} {\bf 60},
922 (1999).

\bibitem{traptree} J.-P. Bouchaud and D.S. Dean, {\it J. Phys. I
(France) } {\bf
5}, 265 (1995); M. Sasaki and K. Nemoto, {\it J. Phys. Soc. Jpn.} {\bf
69} Suppl. A, 238 (2000).

\bibitem{domains} D. S. Fisher and D. A. Huse, {\it Phys. Rev. B} {\bf
38}, 373 and 386 (1988); G. J. M Koper and H. J. Hilhorst, {\it J.
Phys.  France} {\bf 49}, 429 (1988).


\bibitem{Review} For a review of different theoretical models leading to
aging, in particular mean-field models, see: J.-P. Bouchaud, L. F.
Cugliandolo, J. Kurchan, M. M\'ezard, in `Spin-glasses and Random
Fields', pp.161-224, A. P. Young edt. (World Scientific, 1998).

\bibitem{numrecent}H. Rieger, {\it Ann. Rev. of Comp. Phys. II},
ed. D. Stauffer (World Scientific, Singapore, 1995); 
E. Marinari, G. Parisi, F.  Ritort and
J.J. Ruiz-Lorenzo, {\it Phys. Rev. Lett.} {\bf 76}, 843 (1996);
E. Marinari, G. Parisi, F. Ricci-Tersenghi and J.J. Ruiz-Lorenzo, {\it
J. Phys. A} {\bf 33}, 2373 (2000); A. Billoire and E. Marinari, {\it
J. Phys. A} {\bf 33}, L265 (2000); 
M. Picco,
F. Ricci-Tersenghi and F. Ritort, {\it Phys. Rev. B} {\bf 63}, 174412
(2001), and preprint cond-mat/0102248; 
T. Komori, H. Yoshino and H.
Takayama, {\it J. Phys. Soc. Jpn.} {\bf 68}, 3387 (1999), {\it J.
Phys. Soc. Jpn.} {\bf 69}, 1192 (2000), {\it J. Phys. Soc. Jpn.} {\bf
69} Suppl. A, 228 (2000). 

\bibitem{memchaos} K. Jonason, E. Vincent, J. Hammann, J.-P.  Bouchaud
and P. Nordblad, {\it Phys. Rev. Lett.} {\bf 81}, 3243 (1998); K.
Jonason, P. Nordblad, E. Vincent, J. Hammann and J.-P. Bouchaud,
{\it Europ. Phys. Jour. B} {\bf 13}, 99 (2000).

\bibitem{Miyashita} S. Miyashita and E. Vincent, to appear in {\it
Eur. Phys. J. B} (2001), e-print cond-mat/0102077.

\bibitem{bray} A.J. Bray and M.A. Moore, {\it Phys. Rev. Lett.} {\bf
58}, 57 (1987).

\bibitem{Yoshino} H. Yoshino, A. Lema\^itre and J.-P. Bouchaud, {\it
Eur. Phys. J. B} {\bf 20}, 367 (2001).

\bibitem{hierarki} Ph. Refregier, E. Vincent, J. Hammann and M. Ocio,
{\it J. Phys. France} {\bf 48}, 1533 (1987); E. Vincent, J.-P.
Bouchaud, J. Hammann and F. Lefloch, {\it Phil. Mag. B} {\bf 71},
489 (1995).

\bibitem{barriergrowth} J. Hammann, M. Lederman, M. Ocio, R.
Orbach and E. Vincent, {\it Physica A} {\bf 185}, 278 (1992).

\bibitem{FMTOgeneral} A. Ito, E. Torikai, S. Morimoto, H. Aruga,
M. Kikuchi, Y. Syono and H. Takei, {\it J. Phys. Soc. Jpn.} {\bf
59}, 829 (1990); K. Gunnarsson, P. Svedlindh, P. Nordblad, L.
Lundgren, H. Aruga and A. Ito, {\it Phys. Rev. B} {\bf 43}, 8199
(1991); A. Ito, H. Aruga, E. Torikai, M. Kikuchi, Y. Syono
and H. Takei, {\it Phys. Rev. Lett.} {\bf 57}, 483 (1986); H. Aruga,
T. Tokoro and A. Ito, {\it J. Phys. Soc. Jpn.} {\bf 57}, 261 (1988).

\bibitem{FMTOaging}  P. Svedlindh, K. Gunnarsson, P. Nordblad,
L. Lundgren, A. Ito and H.
Aruga, {\it J. Magn. Magn. Mat.} {\bf 71}, 22 (1987); P.
Svedlindh, K. Gunnarsson, J-O. Andersson, H. Aruga Katori and A.
Ito, {\it Phys. Rev. B} {\bf 46}, 13867 (1992).

\bibitem{Nordblad} C. Djurberg, K. Jonason and P. Nordblad, {\it
Eur. Phys. J. B} {\bf 10}, 15 (1999).

\bibitem{foot1}The ferromagnetic transition of the non-diluted
$CdCr_{2}S_4$ is a textbook example of 3d Heisenberg transition. Spin
wave dispersion curves confirm the very weak anisotropy
\cite{Pouget}. On the other hand, in $CdCr_{1.7}In_{0.3}S_4$ (as well
as in other examples), the shape of the irreversibility lines compares
well with the mean-field transition lines for the Heisenberg spin
glass with weak anisotropy (see refs. in \cite{Lefloch}).

\bibitem{Lefloch} F. Lefloch, J. Hammann, M. Ocio and E. Vincent,
{\it Physica B} {\bf 204}, 63 (1994).


\bibitem{Pouget} S. Pouget, M. Alba, M. Fanjat and M. Nogu\`es,
{\it Physica B}, {\bf 180-181}, 244 (1992); S. Pouget and M. Alba,
{\it J. Phys. Condens. Matter}, {\bf 7}, 4739 (1995).

\bibitem{Nogues} M. Alba, J. Hammann and M. Nogu\`es, {\it J.
Phys. C}, {\bf 15}, 5441 (1982); E. Vincent, J. Hammann and M.
Alba, {\it Solid State Comm.} {\bf 58}, 57 (1986).

\bibitem{chisecpowerlaw} M. Alba, J. Hammann, M. Ocio, Ph.
Refregier and H. Bouchiat, {\it J. Appl. Phys.} {\bf 61}, 3683
(1987).

\bibitem{xiOrbach} Y. G. Joh, R. Orbach, G. G. Wood, J. Hammann
and E. Vincent, {\it Phys. Rev. Lett.} {\bf 82}, 438 (1999).

\bibitem{bouchaudNew} J.-P. Bouchaud, V. Dupuis, J. Hammann and
E. Vincent, e-print cond-mat/0106539.

\bibitem{Helene} H. Bouchiat, {\it J. Physique (France)} {\bf 47}, 71 (1986).

\bibitem{Tholence} J. Souletie and J.L. Tholence, {\it Phys. Rev. B
Rapid Comm.} {\bf 32}, 516 (1985).

\bibitem{Isingpreprint} New {\it dc} experiments have been simultaneously
reported on another Ising sample, which show the existence of rejuvenation and memory effects in
agreement with our present {\it ac} results; see R. Mathieu,
P. E. J\"onsson, P. Nordblad, H. Aruga Katori and A. Ito, preprint cond-mat/0104333.

\bibitem{Balents} L. Balents, J.-P. Bouchaud and M. M\'ezard, {\it
J. Phys. I France} {\bf 6}, 1007 (1996); J.-P. Bouchaud in {\it
Soft and Fragile Matter}, ed. M. E. Cates and M. R. Evans
(Institute of Physics Publishing, Bristol and Philadelphia, 2000)
(preprint cond-mat/9910387).

\bibitem{Meyer} K.A. Meyer and M.B. Weissman, {\it Phys. Rev. B} {\bf
51}, 8221 (1995); M.B. Weissman, N.E. Israeloff and G.B. Alers, {\it
J.M.M.M.} {\bf 144}, 87 (1992).

\bibitem{Campbell} I. Campbell, D. Petit, P.-O. Mari and L. Bernardi,
{\it J. Phys. Soc. Jpn.} {\bf 69} Suppl. A, 186 (2000); D. Petit and
I. Campbell, unpublished.

\end{thebibliography}
\end{document}